\documentclass[reprint, amsmath,amssymb, aps]{revtex4-2}
\usepackage{graphicx}
\usepackage[export]{adjustbox}
\usepackage{physics}
\usepackage{dcolumn}
\usepackage{bm}
\usepackage{appendix}

\begin{document}

\preprint{APS/123-QED}

\title{General theory of a moving Fabry-Perot interferometer and its application to the Pound-Drever-Hall technique}

\author{Lingze Duan}
\email{lingze.duan@uah.edu}
\affiliation{Department of Physics and Astronomy, The University of Alabama in Huntsville, Huntsville, AL 35899, USA.}

\date{\today}

\begin{abstract}
In a follow-up effort to our prior report on the optical transmission of a moving Fabry-Perot interferometer \cite{Pyvovar}, this work seeks to establish a general framework that describes the transmission and reflection properties of a Fabry-Perot interferometer when key components in its operation system, including the light source, the detector and the interferometer itself, have relative motions against each other along their common optical axis. Our analysis indicates that these movements result in various new factors in the transmission and reflection coefficients, which all find their roots in the Doppler effect. As a demonstration of its potential application, the new theory is applied to the Pound-Drever-Hall frequency-locking technique. It is shown that velocity-induced frequency modulations are effectively added to the laser frequency due to the motions, and such excess frequency noise can be impactful in certain applications.
\end{abstract}

\maketitle


\section{Introduction}

The Fabry-Perot interferometer (FPI) is a versatile optical device widely used in precision measurement, sensing, and spectroscopy \cite{Vaughan}. It is capable of achieving very high wavelength resolutions through the phenomenon of multiple beam interference between two parallel reflective surfaces. Over the course of the past century, considerable effort has been devoted to expanding the application of the basic concept of FPI, whereas little attention has been given to generalizing the concept itself, especially with regard to the effect of motion.   

The impact of motion on the optical properties of the FPI becomes a relevant question in recent years largely due to the emergence of hybrid interferometers, where an FPI is often nested in a host interferometer to serve as a multiplier of the optical path \cite{Abbott,Graf_et_al,Khalil,Duan}. Since the host interferometer is very sensitive to the transmission phase of the FPI, any phase shift induced by the motion of the FPI would contribute to the output of the hybrid interferometer. Meanwhile, advances in laser and optical waveguide technologies have made it technologically feasible to reliably interrogate a moving FPI. It is under this context that a generalization of the conventional theory of the FPI to include the effects of motion becomes necessary.

In our previous report \cite{Pyvovar}, we have pointed out that a uniform motion of the FPI along its optical axis in the lab frame causes a rescaling of the round-trip phase in the transmission coefficient, which intimately depends on the velocity. In the current work, we expand this concept to allow both the light source and the FPI to independently move. We also investigate the reflection of the FPI in addition to the transmission. As such, we are able to establish a generalized theory for the FPI, encompassing all possible effects induced by the movements of both the interrogation source and the FPI itself.

Before delving into the theory, clarification needs to be made regarding the general concept. There is a common misconception that the effects caused by a moving FPI are simply due to the relative motion between the FPI and the interrogation light source. In other words, the changes to the FPI response reported in \cite{Pyvovar} should also be observed by staying in the co-moving frame of the FPI (i.e., keeping the FPI stationary) while letting the light source move toward the opposite direction. As will be shown in the following discussion, this view is erroneous as it fails to account for an important aspect in the overall observation scheme: the state of motion of the observer itself. In fact, the relative motion between the FPI and the observer and the relative motion between the FPI and the light source have different impacts on the measured FPI response. The phase rescaling factor reported in \cite{Pyvovar} can never be replicated by an observer remaining stationary to the FPI; it is a fundamentally new effect that appears to have been overlooked in the past. 

In the following sections, we will first develop a set of generalized formulations that describe the transmission and the reflection properties of the FPI under different motion scenarios. Then, we will explore the potential applications of the new theory, in particular, the impacts of motions on laser frequency locking based on the Pound-Drever-Hall (PDH) technique. 


\section{A Generalized Theory}

\begin{figure}[ht]
\centering
\includegraphics[width=0.95\linewidth]{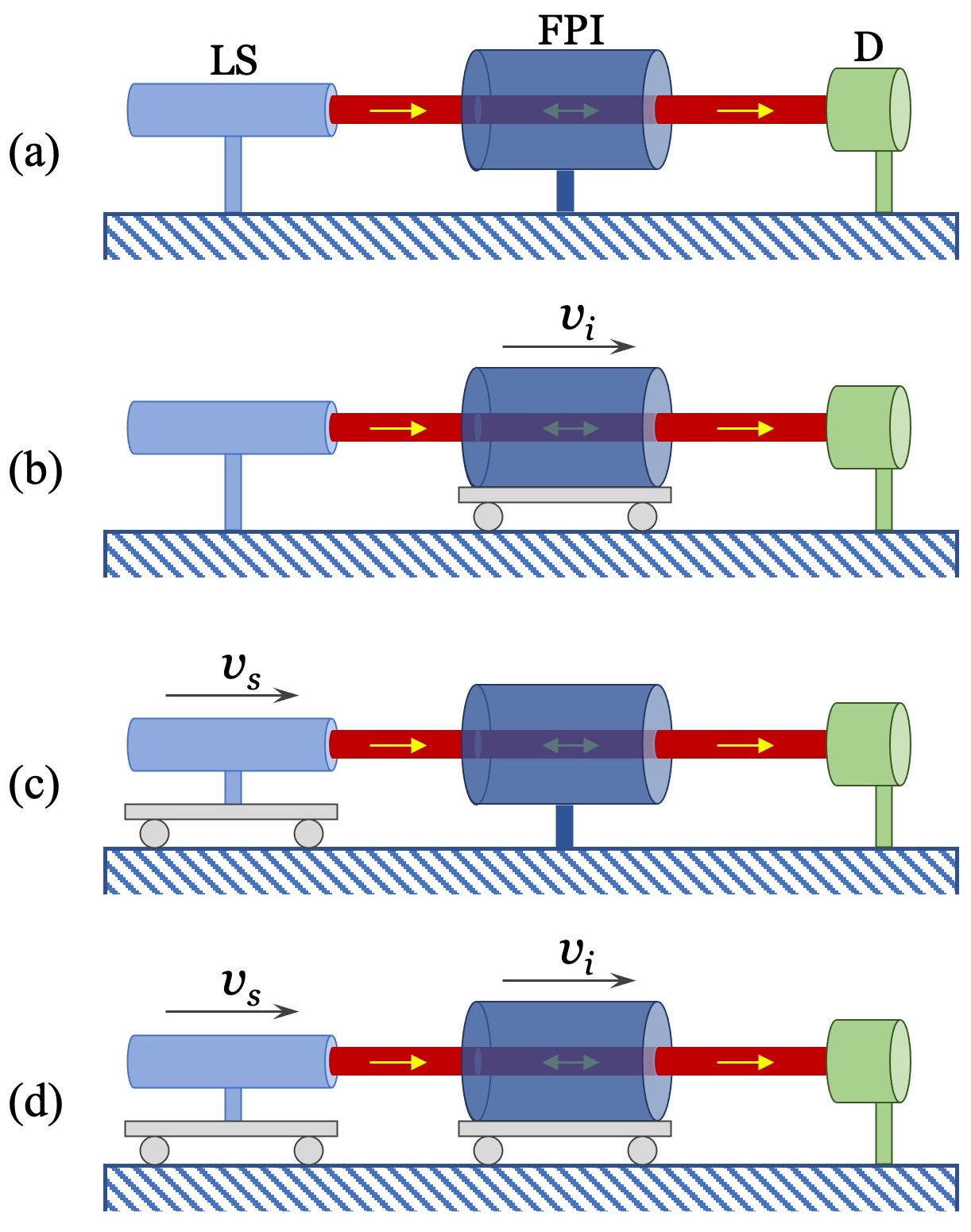}
\caption{The three key components in the operation system of a passive FPI and the four motion schemes ((a)\textendash(d)) studied in this work. LS: light source, D: detector.}
\label{fig:comparison} 
\end{figure}

Let us first define the system under study. In this work, we focus on passive FPIs, whose operation involves three key components: a light source, an FPI, and a detector, as conceptually shown in Fig.~\ref{fig:comparison}(a). All three components can in principle move independently along the optical axis. Since the action of \textquotedblleft observation\textquotedblright is effectively performed by the detector in the current context, without losing any generality, we consider the detector as the \textquotedblleft observer\textquotedblright and define its frame as the \textquotedblleft lab frame\textquotedblright. The states of motion of the other two components in the lab frame create four scenarios, depicted here in Fig.~\ref{fig:comparison}(a)\textendash(d): (a) both the light source and the FPI are stationary; (b) the light source is stationary while the FPI is moving; (c) the light source is moving while the FPI is stationary; (d) both the light source and the FPI are moving. Note that these are four distinct scenarios, which cannot be duplicated by another one as discussed in the following.

Here are some general conditions we follow throughout this paper. The FPI is treated as a rigid body without deformation. All the motions are considered uniform, i.e., at constant velocities. This generally suffices because, as previously shown \cite{Pyvovar}, nonuniform motions can be fairly well described by a simple generalization of the uniform-motion model within a large acceleration range. Finally, all discussions are confined in the nonrelativistic regime given the negligible impact of relativistic effects for most applications \cite{Pyvovar}.

\subsection{Stationary Light Source and Stationary FPI}

We begin with the simplest case, where both the light source and the FPI are fixed in the lab frame and hence remain stationary to the detector (i.e., the observer) as shown in Fig.~\ref{fig:comparison}(a). This is the conventional way of operating the FPI. We can directly write out the transmission and the reflection coefficients of the FPI as
\begin{equation}
    T \equiv \frac{\mathcal{E}_T}{\mathcal{E}_I} = \frac{(1-r^2)e^{-inkd}}{1-r^2e^{-2inkd}},
    \label{T_Case_1}
\end{equation}
and
\begin{equation}
    R \equiv \frac{\mathcal{E}_R}{\mathcal{E}_I} = \frac{r(1-e^{-2inkd})}{1-r^2e^{-2inkd}},
    \label{R_Case_1}
\end{equation}
where $\mathcal{E}_I$, $\mathcal{E}_R$ and $\mathcal{E}_T$ are the electric fields of the incident, reflected and transmitted waves, respectively, and $k$ is the wave number in vacuum. The reflection coefficients of the two mirrors are assumed to be identical here and are denoted as $r$. With a proper length $d$, the FPI is considered to be made of a homogeneous, isotropic, and lossless optical medium of refractive index $n$. The corresponding FPI transmittance and reflectance are given by 
\begin{equation}
    |T|^2=\frac{1}{1 + (2\mathcal{F}/\pi)^2 \sin^2{(nkd)}},
    \label{Transmittance_Case_1}
\end{equation}
and 
\begin{equation}
    |R|^2=\frac{(2\mathcal{F}/\pi)^2 \sin^2{(nkd)}}{1 + (2\mathcal{F}/\pi)^2 \sin^2{(nkd)}},
    \label{Reflectance_Case_1}
\end{equation}
where $\mathcal{F}=\pi r/(1 - r^2)$ is the finesse of the FPI.


\subsection{Stationary Light Source and Moving FPI}

Let us move on to the second case, where the light source is fixed in the lab frame while the FPI is traveling uniformly along the optical axis, as shown in Fig.~\ref{fig:comparison}(b). We have previously analyzed this case for optical transmission and have shown that a velocity-dependent scaling factor needs to be added to the round-trip phase to account for the effect of uniform motion \cite{Pyvovar}. The resulting transmission coefficient can be written as 
\begin{equation}
    T = \frac{(1-r^2)e^{-i\zeta nkd}}{1-r^2e^{-2i\zeta nkd}},
    \label{T_Case_2}
\end{equation}
where $\zeta$ is defined as
\begin{equation}
\zeta = \sqrt{\frac{c-v_i}{c+v_i}},
\label{zeta}
\end{equation}
with $v_i$ and $c$ being the velocity of the interferometer and the speed of light in vacuum, respectively.

\begin{figure}[ht]
\centering
\includegraphics[width=0.95\linewidth]{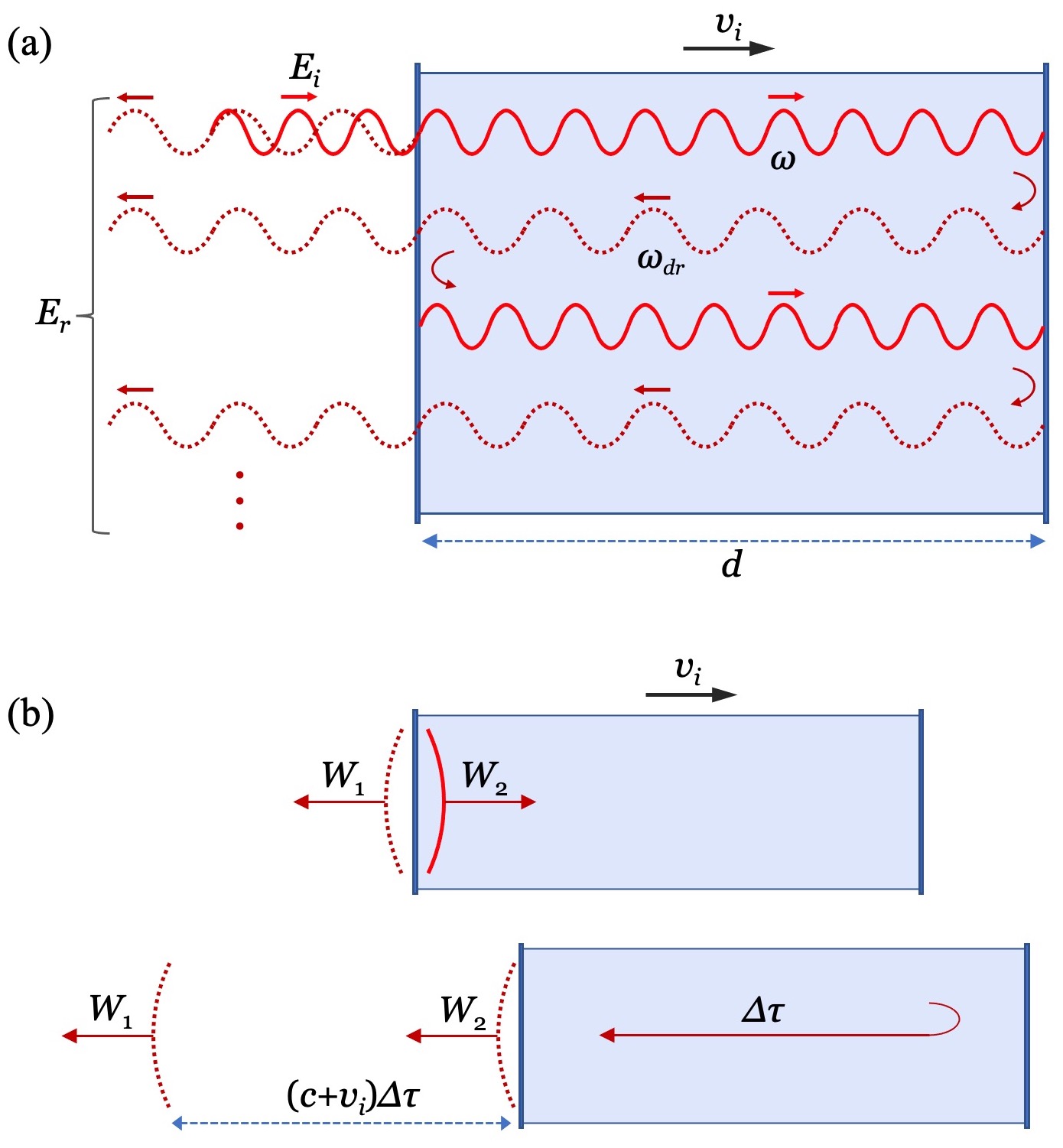}
\caption{(a) The concept of multiple-wave superposition for a uniformly moving FPI. (b) Finding the round-trip phase delay by following the propagation of a wavefront.}
\label{fig:multiwave}
\end{figure}

To find the reflection coefficient of the FPI, we follow the same strategy as previously used to derive the transmission coefficient and construct the superposition of consecutively reflected wavefronts \cite{Pyvovar}. Fig.~\ref{fig:multiwave} conceptually illustrates this process. First, we note that, unlike transmission, waves reflected off a uniformly moving FPI have a different frequency from the incident wave due to the Doppler effect. At any moment of time, if the incident field on the input plane of the FPI is denoted as $\mathcal{E}_I = E_i e^{i\omega t}$, and the total reflected field on the same plane is written as $\mathcal{E}_R = E_r e^{i\omega_{dr} t}$, then $\omega$ and $\omega_{dr}$ are related through the relation \cite{Ives}
\begin{equation}
\omega_{dr} = \omega \frac{c-v_i}{c+v_i},
\label{Doppler}
\end{equation}
where $v_i$ is positive when the FPI is moving along the same direction as the incident wave. $E_i$ and $E_r$ are complex amplitudes for the incident and the reflected fields, respectively. When the FPI is moving uniformly, there is a steady relation between $E_r$ and $E_i$, which can be derived through multiple-wave superposition as shown in Fig.~\ref{fig:multiwave}(a). It is straightforward to verify that such an analysis results in a general relation 
\begin{equation}
    E_r = \frac{r(1-e^{-i\delta})}{1-r^2e^{-i\delta}} E_i,
    \label{E_r_and_E_i}
\end{equation}
where $\delta$ is the round-trip phase delay in a uniformly moving FPI. To find $\delta$, we follow two consecutive wavefronts $W_1$ and $W_2$ through a round trip, as illustrated in Fig.~\ref{fig:multiwave}(b). Their phase delay is given by 
\begin{equation}
    \delta = \frac{\omega_{dr}}{c} (c+v_i) \Delta \tau,
    \label{Delta_Case_2}
\end{equation}
where $\Delta \tau$ is the round-trip time of the wavefront $W_2$ inside the FPI. Previously \cite{Pyvovar}, we have shown that $\Delta \tau$ is given by,
\begin{equation}
\Delta \tau = \frac{2nd}{\sqrt{c^2-v_i^2}}.
\label{Delta_tau}
\end{equation}
Applying \eqref{Doppler} and \eqref{Delta_tau} to \eqref{Delta_Case_2} leads to 
\begin{equation}
\delta = 2\zeta nkd,
\label{delta}
\end{equation}
which is the same round-trip phase used in the transmission coefficient \eqref{T_Case_2} with $\zeta$ given by \eqref{zeta}.

Combining \eqref{E_r_and_E_i} and \eqref{delta} yields the total reflected field
\begin{equation}
    \mathcal{E}_R = \frac{r(1-e^{-2i\zeta nkd})}{1-r^2e^{-2i\zeta nkd}} E_i e^{i\omega_{dr} t}.
    \label{mathcal_E_R}
\end{equation}
Taking the ratio of $\mathcal{E}_R$ and $\mathcal{E}_I$ results in the reflection coefficient of a uniformly moving FPI
\begin{equation}
    R = \frac{r(1-e^{-2i\zeta nkd})}{1-r^2e^{-2i\zeta nkd}} e^{-2i \left(\frac{v_i}{c}\right) \omega t},
    \label{R_Case_2}
\end{equation}
where the nonrelativistic condition $v_i \ll c$ has been used to simplify the Doppler-shift term. This coefficient \eqref{R_Case_2} clearly shows the impact of motion to the reflected field in two aspects: \textit{i)} a velocity-dependent scaling factor $\zeta$ in the round-trip phase (same as the transmission scaling factor), and \textit{ii)} a Doppler frequency shift proportional to twice the velocity of the FPI, which does not exist in the transmitted field.


\subsection{Moving Light Source and Stationary FPI}

The next case to be considered is the scenario where the FPI is fixed in the lab frame while the light source is uniformly moving along the optical axis as shown in Fig.~\ref{fig:comparison}(c). Since the FPI is stationary relative to the detector, from the observer point of view, its behavior is same as in Case A. The moving light source, however, introduces a Doppler shift that has to be factored in when considering the round-trip phase inside the FPI. With this basic understanding, we can utilize the relations \eqref{T_Case_1} and \eqref{R_Case_1} to construct the transmission coefficient and the reflection coefficient of the FPI as 
\begin{equation}
    T = \frac{(1-r^2)e^{-ink_d d}}{1-r^2e^{-2ink_d d}},
    \label{T_Case_3}
\end{equation}
and
\begin{equation}
    R = \frac{r(1-e^{-2ink_d d})}{1-r^2e^{-2ink_d d}},
    \label{R_Case_3}
\end{equation}
where $k_d$ is the wave number after the Doppler shift. Given the Doppler-shifted wavelength $\lambda_d = \lambda (1-v_s/c)$, where $\lambda$ is the actual wavelength of the light source and $v_s$ is the velocity of the light source (positive when moving toward the FPI), it is straightforward to show that $k_d$ can be written as 
\begin{equation}
    k_d = k \left(1 + \frac{v_s}{c}\right),
    \label{k_d}
\end{equation}
where $k = 2\pi/\lambda$ is the wave number when the light source is at rest.


\subsection{Moving Light Source and Moving FPI}

In this last case as shown in Fig.~\ref{fig:comparison}(d), we analyze the most generic scenario, where both the light source and the FPI move independently along the optical axis in the lab frame, at $v_s$ and $v_i$, respectively. Based on the discussions in the previous sections, we can qualitatively summarize the impacts of motion in the operation of an FPI as: a nonzero velocity of the FPI introduces a scaling factor $\zeta$ in the round-trip phase, while a nonzero velocity of the light source changes the wave number through the Doppler-shifted wavelength. 

Since these two processes are independent from each other, it is conceivable that, when they are present at the same time, the overall response of the FPI is a simple combination of the two effects. In the case of transmission, combining \eqref{T_Case_2} and \eqref{T_Case_3} leads to
\begin{equation}
    T = \frac{(1-r^2)e^{-i\zeta nk_d d}}{1-r^2e^{-2i\zeta nk_d d}},
    \label{T_Case_4}
\end{equation}
where $\zeta$ and $k_d$ are defined by \eqref{zeta} and \eqref{k_d}, respectively. For reflection, a similar treatment to \eqref{R_Case_2} and \eqref{R_Case_3} yields
\begin{equation}
    R = \frac{r(1-e^{-2i\zeta nk_d d})}{1-r^2e^{-2i\zeta nk_d d}} e^{-2i \left(\frac{v_i}{c}\right) \omega t}.
    \label{R_Case_4}
\end{equation}

Finally, we summarize the latter three cases by giving the general form of the transmittance and the reflectance of the FPI when motion is considered. Comparing \eqref{T_Case_4} and \eqref{R_Case_4} with \eqref{T_Case_1} and \eqref{R_Case_1}, it immediately becomes clear that a straightforward generalization of \eqref{Transmittance_Case_1} and \eqref{Reflectance_Case_1} leads to
\begin{equation}
    |T|^2=\frac{1}{1 + (2\mathcal{F}/\pi)^2 \sin^2{(\zeta nk_d d)}},
    \label{Transmittance_Case_4}
\end{equation}
and 
\begin{equation}
    |R|^2=\frac{(2\mathcal{F}/\pi)^2 \sin^2{(\zeta nk_d d)}}{1 + (2\mathcal{F}/\pi)^2 \sin^2{(\zeta nk_d d)}}.
    \label{Reflectance_Case_4}
\end{equation}


\section{Application in the PDH technique}

To demonstrate the application of the general theory, let us evaluate the impact of motions in a Pound-Drever-Hall (PDH) frequency locking system \cite{Drever}. The PDH technique has been widely used in laser frequency stabilization \cite{Salomon,Gibble}, precision metrology \cite{McNamara,Lawall}, and optical sensing \cite{Zhan,Hoque}. Its operation involves comparing the laser frequency with a resonant frequency of a reference FPI and subsequently generating an error signal proportional to this frequency difference \cite{Black}. The error signal is then fed into a feedback loop to drive the control system.

\begin{figure}[ht]
\centering
\includegraphics[width=0.95\linewidth]{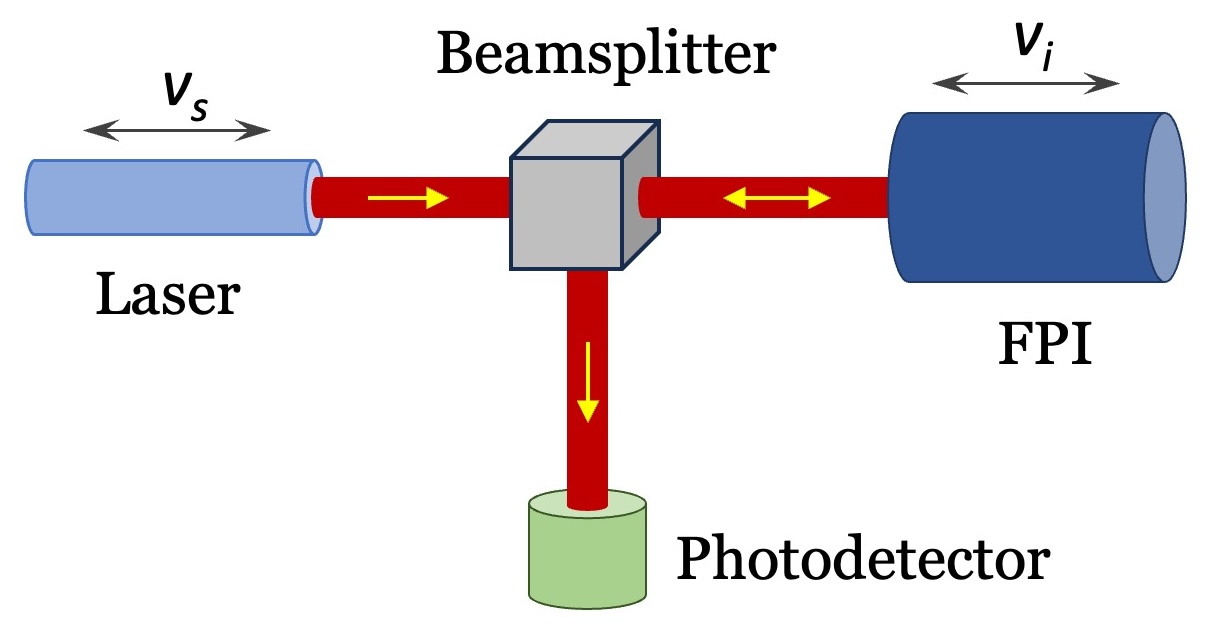}
\caption{Simplified optical layout for a typical Pound-Drever-Hall frequency locking system. In the current discussion, the laser and the FPI are considered movable in the longitudinal direction.}
\label{fig:PDH} 
\end{figure}

For the purpose of the current analysis, we neglect the ancillary components of the PDH system such as the electro-optic modulator and all the electronics, keeping only the core optical components: a laser, an FPI, a beamsplitter and a photodetector. A conceptual layout of the scheme is shown in Fig.~\ref{fig:PDH}. We further assume the laser and the FPI can move along their optical axes while the beamsplitter and the photodetector remain stationary in the lab frame.

Let us first examine the case with a moving laser and a fixed FPI (see Fig.~\ref{fig:comparison}(c)). The reflection coefficient of the FPI is given by \eqref{R_Case_3}. Following the discussion and the notations in \cite{Black} while incorporating \eqref{k_d}, we can rewrite the round-trip phase into
\begin{equation}
\delta = 2nk_dd = \frac{\omega + \delta \omega_s}{\Delta \nu_{fsr}},
\label{RT_Phase}
\end{equation}
where $\omega$ is the original frequency of the laser, i.e., the frequency in its rest frame, $\delta \omega_s = \omega(v_s/c)$ is the Doppler shift caused by the laser movement, and $\Delta \nu_{fsr} = c/(2nd)$ is the free spectral range of the FPI. Equation \eqref{RT_Phase} suggests that a longitudinal motion of the laser effectively introduces a frequency drift to the laser, which is proportional to the velocity of the movement. For a PDH locking system, such a frequency drift is indistinguishable from an intrinsic laser frequency change. In other words, the motion effectively adds another frequency noise to the laser.

Next, we analyze the case where the laser is fixed while the FPI is moving (see Fig.~\ref{fig:comparison}(b)). This scenario is more complex because two factors are involved in the process: an oscillation term representing a Doppler shift of $-2(v_i/c)\omega$ and a rescaling factor $\zeta$ in the round-trip phase, as pointed out earlier in Section II. B. 

First, let us focus on the reflection-induced Doppler term $\exp{[-2i(v_i/c)\omega t]}$. It introduces a universal frequency shift to any wave reflected by the FPI. In the PDH scheme, the wave incident on the FPI is a frequency-modulated (FM) optical carrier \cite{Drever, Black}. It has been shown that a periodically modulated wave reflected off a uniformly moving target remains as a periodically modulated wave, but with a Doppler-shifted modulation frequency \cite{Sobolev,Sung}. Such a feature can be easily seen in the current context by inspecting how the optical carrier and its two FM sidebands change frequencies upon reflecting from the FPI. If we denote the frequency of the laser as $\omega$ and the frequencies of the FM sidebands as $\omega+\Omega$ and $\omega-\Omega$, where $\Omega$ is the modulation frequency of the electro-optic modulator, then according to \eqref{R_Case_2}, the three frequencies are shifted to
\begin{equation}
\begin{cases}
    \omega \rightarrow \omega+K\omega = (1+K)\omega,\\
    \omega+\Omega \rightarrow (\omega+\Omega)+K(\omega+\Omega) = (1+K)\omega+(1+K)\Omega,\\
    \omega-\Omega \rightarrow (\omega-\Omega)+K(\omega-\Omega) = (1+K)\omega-(1+K)\Omega,
\end{cases}
\label{PDH_f_shift}
\end{equation}
where $K=-2(v_i/c)$. It is evident from \eqref{PDH_f_shift} that the reflected light off a moving FPI contains a Doppler-shifted optical carrier $(1+K)\omega$ and a Doppler-shifted FM frequency $(1+K)\Omega$. Since the PDH technique is insensitive to the carrier frequency \cite{Black}, the impact of the Doppler term in \eqref{R_Case_2} is mainly contributed by the FM frequency shift. In principle, such a change can introduce an additional noise to the PDH error signal through the subsequent homodyne process. Practically, the severity of this additional noise depends on the top speed of the FPI as well as the noise budget of the locking system. As will be shown later, a reasonable speed range for the FPI is on the order of mm/s to cm/s, which leads to a $K$ factor of the order of $10^{-11}-10^{-10}$. The absolute frequency shift in a typical 10-MHz FM frequency is about $0.1-1$ mHz. For most common applications, this level of uncertainty can be well within the tolerable range. 

The second effect caused by a moving FPI is the velocity-dependent scaling factor $\zeta$. It is not hard to see by comparing \eqref{R_Case_2} and \eqref{R_Case_3} that $\zeta$ and $k_d$ play similar roles in the FPI reflection coefficient, viz., adding a Doppler drift to the optical frequency. This becomes more clear when \eqref{zeta} is rewritten as $\zeta \approx (1-v_i/c)$ and substituted into the round-trip phase given by \eqref{delta}, which yields
\begin{equation}
\delta = \frac{\omega - \delta \omega_i}{\Delta \nu_{fsr}},
\label{RT_Phase_case2}
\end{equation}
where $\delta \omega_i = \omega(v_i/c)$ is an equivalent Doppler drift in the optical frequency. Note that this result is consistent with our prior report, where we concluded that a nonzero velocity of the FPI would rescale its resonance peaks by a factor of $(1+v_i/c)$ \cite{Pyvovar}.

Finally, we discuss the general case where both the laser and the FPI are moving (see Fig.~\ref{fig:comparison}(d)). A quick comparison of the general reflection coefficient \eqref{R_Case_4} with the previous two cases \eqref{R_Case_2} and \eqref{R_Case_3} indicates that the overall impacts of these motions on a PDH locking system combine all the effects pointed out earlier. Of particular interest here is the overall round-trip phase, where two separate Doppler drifts must be factored in. Combining \eqref{delta}, \eqref{RT_Phase} and \eqref{RT_Phase_case2}, it is straightforward to show that the general round-trip phase is given by
\begin{equation}
\delta = \frac{\omega}{\Delta \nu_{fsr}}\left( 1-\frac{v_i}{c}\right)\left( 1+\frac{v_s}{c}\right) \approx \frac{\omega - \delta \omega_i + \delta \omega_s}{\Delta \nu_{fsr}},
\label{RT_Phase_case3}
\end{equation}
where the second-order term has been neglected. Evidently, the motions of both the laser and the FPI contribute to the overall frequency drift. The sign difference between $\delta \omega_i$ and $\delta \omega_s$ is simply due to the different definitions of positive $v_i$ and $v_s$. 

An interesting observation is worth being pointed out here. When $v_i$ and $v_s$ are equal and pointing toward the same direction, the $\delta \omega_i$ and $\delta \omega_s$ terms in \eqref{RT_Phase_case3} cancel out (only up to the first order), leaving $\delta \approx \omega/\Delta \nu_{fsr} = 2nkd$ taking a form as if the laser and the FPI are both stationary. This suggests that, by keeping the laser and the FPI relatively stationary to each other, the impact of velocity can be mitigated. In other words, a relative motion between the detector and the laser-FPI pair is far less influential to the PDH error signal than a relative motion between the detector and either the laser or the FPI individually.

In reality, the PDH technique is rarely used to lock uniformly moving lasers or FPIs. A more realistic scenario is the mechanical vibration of optical fixtures. In any optical systems, the mechanical structures these optical components are mounted on inevitably suffer from spontaneous vibrations. The impacts of such vibrations have long been analyzed from the displacement point of view but rarely from the velocity point of view. An in-depth discussion of the effect of vibration velocity on PDH locking is beyond the scope of this report. Here, we opt to use a simple example to loosely gauge the potential impact.

Suppose one of the optical components (laser or FPI) is mounted on a mechanical structure that vibrates in the direction of the laser beam at an intrinsic resonance frequency of 1 kHz, with a peak-to-peak amplitude of 1 $\mu$m. Vibrations of such a scale are quite common in optical structures \cite{Forward,Nachman}. According to \eqref{RT_Phase_case3}, a frequency modulation at 1 kHz is effectively added to the laser frequency. To evaluate the scale of this additional frequency noise, we use the peak-to-peak velocity swing $\Delta v_{pp}$ as an indicator, and $\Delta v_{pp} \sim 6$ mm/s in the current case, which leads to a Doppler ratio $\Delta v_{pp}/c \sim 2 \times 10^{-11}$. For a typical laser frequency at 400 THz (750 nm), the corresponding frequency swing is about 8 kHz. This level of excess frequency noise is nonnegligible in high-precision PDH applications \cite{Zhang}. The example thus demonstrates the necessity to take into account potential velocity-induced frequency noises in laser frequency locking based on the PDH technique.

 \section{Conclusion}
In conclusion, we have formulated a general theory describing the transmission and reflection behaviors of the FPI when key components in its operation system are moving relative to each other along their common optical axis. It has been found that a relative motion between the light source and the detector can be accounted for by a Doppler-shifted wave number (or frequency) in the transmission and reflections coefficients, whereas a relative motion between the FPI and the detector introduces two effects in the reflection of the FPI: a Doppler term that shifts the frequency of the incident wave by $-2(v_i/c)$ and a rescaling factor $\zeta$ in the round-trip phase. To showcase the applications of this new theory, we have investigated velocity-induced excess noise in PDH laser frequency locking. Our analysis shows that tiny mechanical vibrations of the laser and the FPI can effectively introduce additional FM noise to the laser frequency, which should be taken into account in high-precision laser frequency stabilization. Overall, it is our hope that this work helps bring necessary attention to the overlooked aspect of velocity-induced effects in metrological systems involving the FPI.

\nocite{*}
\bibliography{apsbib}

\end{document}